\documentclass[aps,twocolumn,showpacs]{revtex4}
\usepackage{graphicx}
\usepackage{bm}
\usepackage{amsmath}
\renewcommand{\a}{\alpha}
\renewcommand{\b}{\beta}
\newcommand{\g}{\gamma}
\renewcommand{\d}{\delta}
\newcommand{\cR}{{\cal R}_{\g\d}}
\newcommand{\cQ}{{\cal Q}_{\g\d}}

\newcommand{\om}{{\omega}}

\begin{document}

\title{Strong Coupling Solver for the Quantum Impurity Model }
\author{Xi Dai$^a$, Kristjan Haule$^b$, Gabriel Kotliar$^c$}
\affiliation{$^a$Department of Physics, University of Hong
Kong,Hong Kong, China} \email[E-mail: ]{daix@hkucc.hku.hk}
\affiliation{$^b$Jo\v zef Stefan Institute, SI-1000 Ljubljana,
Slovenia} \affiliation{$^c$Department of Physics and Center for
Material Theory, Rutgers University, Piscataway, NJ 08854, USA}
\date{\today}

\begin{abstract}

We propose a fast impurity solver for the general quantum impurity
model based on the perturbation theory around the atomic limit,
which can be used in combination with the local density
approximation (LDA) and the dynamical mean field theory (DMFT). We
benchmark the solver in the two band Hubbard model within DMFT
against quantum Monte Carlo (QMC) and numerical renormalization
group (NRG) results. We find that the solver works very well in
the paramagnetic Mott insulator phase. We also apply this impurity
solver to the DMFT study of the anti-ferromagnetic phase
transition in the unfrustrated Bethe lattice. The Neel temperature
obtained by the fast impurity solver agrees very well with the QMC
results in the large Hubbard U limit.  The method is a promising
tool to be used in combination with the LDA+DMFT to study Mott
insulators starting  from  first principles.

\end{abstract}

\pacs{71.10Fd,71.20.Be}

\maketitle

\section{Introduction}

\label{Intro} Recently, much effort has been devoted to develop
methods for {\it ab initio} investigations of real materials with
strongly correlated electrons. A  most promising tool was build by
combining the conventional first principle methods, such as the
density functional theory in the  local density approximation
(LDA), with the newly developed dynamical mean field theory
(DMFT)\cite{DMFT_review}. Many numerical schemes such as
LDA+DMFT\cite{LDA+DMFT}, LDA++\cite{LDA++}, GW+DMFT\cite{GW+DMFT}
has been proposed and applied to various systems\cite{Pu,V2O3}.

The application of DMFT to the real material requires a fast
scheme to solve the generalized Anderson impurity model. Most of
the impurity solvers, such as the iterative perturbation theory
(IPT)\cite{IPT,combine}, the non-crossing approximation
(NCA)\cite{NCA}, the slave boson mean-field\cite{SB}, the equation
of motion method\cite{EOM},  exact diagonalization based methods
\cite{ED}   and quantum Monte Carlo (QMC)\cite{QMC}, have been
developed for simplified multi-orbital Anderson models, usually
assuming SU(N) symmetry. While very few tools are available for
the study of general  Anderson impurity models generated by
realistic DMFT calculations. Therefore it is important to develop
impurity solvers which can be used for a very general case. In the
weak coupling limit when the system is in the metallic phase, the
fluctuation exchange approximation(FLEX)\cite{FLEX} has been
proposed and implemented. On the other hand  for the Mott
insulator phase, when the local interaction term is very strong,
we are  still lack of such a general impurity solver which can
treat models  without SU(N) symmetry and containing general
crystal fields and multiplet terms. Recent studies on the Mott
insulators\cite{Nagaosa}, i.e. $LaMnO_3$, $V_2O_3$, $LaTiO_3$ and
$YTiO_3$\cite{LaTiO3}, discovered a variety of phenomena including
orbital order-disorder transition, charge density wave and
anti-ferromagnetism. Therefore the first principle study on the
Mott insulator material with or without long ranger order becomes
a very important issue for both the theoretical understanding  of
these materials and material designing of these type of compounds.

\section{derivation of the method}

In this paper, we propose an impurity solver which is based on the
perturbation theory around the atomic limit. We will consider the
most general Anderson impurity model generated by the LDA+DMFT
calculation
\begin{equation}
H_{total}=H_{local}+H_{band}++H_{v}
\end{equation}

\begin{eqnarray}
H_{local}&=&\sum_{\a\b} t_{\a\b}f_\a^\dagger f_\b\nonumber\\
&+&\frac{1}{2}\sum_{\a\b\g\d}
U_{\a\b\g\d}f_\a^\dagger f_\b^\dagger f_\d f_\g \\
H_{band}&=&\sum_{k\a}\epsilon_{k\a} c_{k\a}^\dagger c_{k\a}\\
H_{v}&=&\sum_{k\a\b}V_{k\a\b}f^\dagger_{\a}c_{k\b}+h.c.
\label{hyb}
\end{eqnarray}
where $H_{local}$ is a very general atomic Hamiltonian and index
$\a$ denotes the spin and orbital degree of freedom. Further,
$H_{band}$ stands for the conducting band which plays the role of
a fermionic bath in DMFT calculations.

Our first step is to diagonalize the atomic Hamiltonian
$H_{local}$ by the exact diagonalization
\begin{equation}
H_{local}=\sum_{m} E_{m} |m\rangle\langle m|
\end{equation}
which can be done for any atom on modern computers. The
hybridization term then takes the form
\begin{equation}
  H_{v}=\sum_{k\a\b}V_{k\a\b}(F^{\alpha\dagger})_{mm'}\;
  |m\rangle\langle m'| c_{k\beta}+h.c.
\end{equation}
where $(F^{\a})_{mm'}=\langle m|f_\a|m'\rangle$, are the matrix
elements of the operator $f_\a$ in the local eigenbase. The atomic
Green's function can be expressed by the eigenstates in the
following way
\begin{eqnarray}
G_{\a\b}^{(atom)}(i\omega)&=&-\int_{0}^{\beta}
e^{i\omega\tau}\langle
T_{\tau}f_{\a}(\tau)f_{\b}^\dagger(0)\rangle d\tau \label{Gatom}
\\
&=&
\sum_{mm'}\frac{(F^\a)_{mm'}(F^{\b\dagger})_{m'm}(X_m+X_{m'})}{i\omega-E_{m'}+E_m},\nonumber
\end{eqnarray}

where $X_m=e^{-\beta E_m}/Z$ is the probability for the atomic
state $|m\rangle$. The atomic self-energy can then be obtained by
the inversion of a matrix
\begin{equation}
\Sigma^{(atom)}(i\omega)=(i\omega+\mu- \hat{t})^{-1}-
{G^{(atom)}(i\omega)}^{-1} \label{atom}.
\end{equation}

This is the zeroth order self-energy in the expansion around the
atomic limit. If we are able to compute the expansion of the
Green's function in powers of the hybridization, i.e.,
$G=G^{(atom)}+G^{(2)}+O(V^4)$, we could also express the
correction to the self-energy using the Dyson equation
\begin{equation}
G=(i\omega+\mu-\hat{t}-\Sigma-\Delta)^{-1}.
\end{equation}
To the lowest order in hybridization, we have
\begin{equation}
\Sigma=\Sigma^{(atom)}+{G^{(atom)}}^{-1}G^{(2)}{G^{(atom)}}^{-1}
-\Delta + O(V^4).
\label{Sigma_exp}
\end{equation}

The expansion of the Green's function in the hybridization can be
done by linked cluster expansion method following Metzner {\it et
al.} \cite{linked_cluster} or using the auxiliary particle method
\cite{Haule} or by straightforward expansion of the following
functional integral
\begin{equation}
G_{\a\b}(\tau)=\frac{\int D[f^\dagger f]\;
f_\b^\dagger(0)f_\a(\tau)\exp(-S)}{\int D[f^\dagger f]\exp(-S)},
\end{equation}
where
\begin{equation}
S = S_{local} + \sum_{\a\b}\int_0^\beta d\tau \int_0^\beta d\tau'
f_\a^\dagger(\tau)\Delta_{\a\b}(\tau-\tau')f_\b(\tau').
\end{equation}
Expanding to the lowest order in $\Delta$, one obtains two first
order terms: a simple one body term from expanding the denominator
and the complicated two-body term from expanding the nominator.
The correction to the atomic Green's function takes the form
\begin{eqnarray}
G^{(2)}_{\a\b}(i\om) = G^{(atom)}_{\a\b}(i\om)\beta\;
\mathrm{Tr}[G^{(atom)}\Delta] + {G^2}_{\a\b}(i\om) \label{G2_one}
\end{eqnarray}
where the two particle Green's function $G^2$ is
\begin{eqnarray}
G^2_{\a\b}(i\om)&=&\int_0^\b d\tau\int_0^\b d\tau_1\int_0^\b
d\tau_2e^{i\omega\tau}\times \\
&&\sum_{\g\d}\langle T_\tau
f_\a(\tau)f_\b^\dagger(0)f_\g^\dagger(\tau_1)f_\d(\tau_2)\rangle_0
\;\Delta_{\g\d}(\tau_1-\tau_2)\nonumber.
\label{dG}
\end{eqnarray}

and the average $\langle\dots\rangle_0$ is the atomic average,
i.e., $\mathrm{Tr}(\exp(-\beta H_{loc})\dots)/Z$. Inserting the
representation of the electron operator $f_\a =
\sum_{mm'}F^\a_{mm'} |m\rangle\langle m'|$ in the above atomic
average, one arrives at

\begin{widetext}
\begin{eqnarray}
G^{2}_{\a\b}(i\om)&=&T\sum_{i\om'}\Delta_{\g\d}(i\omega') \int
d\tau\int d\tau_1\int d\tau_2 \langle T_\tau
f_\a(\tau)f_\g^{\dagger}(\tau_1)f_\d(\tau_2)f_\b^\dagger(0)
\rangle e^{i\om\tau+i\om'(\tau_2-\tau_1)}\\
&=&\sum_{0123}\frac{e^{-\beta E_0}}{Z}
T\sum_{i\om'}\Delta_{\g\d}(i\omega')
\int d\tau\int d\tau_1\int d\tau_2\Big[\nonumber\\
&& \theta(\tau>\tau_1>\tau_2>0)\
(F^\a)_{01}(F^{\g\dagger})_{12}(F^{\d})_{23}(F^{\b\dagger})_{30}\;
e^{\tau(E_0-E_1+i\om)+\tau_1(E_1-E_2-i\om')+\tau_2(E_2-E_3+i\om')}\nonumber\\
&-& \theta(\tau>\tau_2>\tau_1>0)\;
(F^\a)_{01}(F^{\d})_{12}(F^{\g\dagger})_{23}(F^{\b\dagger})_{30}\;
e^{\tau(E_0-E_1+i\om)+\tau_2(E_1-E_2+i\om')+\tau_1(E_2-E_3-i\om')}\nonumber\\
&-& \theta(\tau_1>\tau>\tau_2>0)\;
(F^{\g\dagger})_{01}(F^{\a})_{12}(F^{\d})_{23}(F^{\b\dagger})_{30}\;
e^{\tau_1(E_0-E_1-i\om')+\tau(E_1-E_2+i\om)+\tau_2(E_2-E_3+i\om')}\nonumber\\
&+& \theta(\tau_2>\tau>\tau_1>0)\;
(F^\d)_{01}(F^{\a})_{12}(F^{\g\dagger})_{23}(F^{\b\dagger})_{30}\;
e^{\tau_2(E_0-E_1+i\om')+\tau(E_1-E_2+i\om)+\tau_1(E_2-E_3-i\om')}\nonumber\\
&+& \theta(\tau_1>\tau_2>\tau>0)\;
(F^{\g\dagger})_{01}(F^{\d})_{12}(F^{\a})_{23}(F^{\b\dagger})_{30}\;
e^{\tau_1(E_0-E_1-i\om')+\tau_2(E_1-E_2+i\om')+\tau(E_2-E_3+i\om)}\nonumber\\
&-& \theta(\tau_2>\tau_1>\tau>0)\;
(F^\d)_{01}(F^{\g\dagger})_{12}(F^{\a})_{23}(F^{\b\dagger})_{30}\;
e^{\tau_2(E_0-E_1+i\om')+\tau_1(E_1-E_2-i\om')+\tau(E_2-E_3+i\om)}\Big]\nonumber
\end{eqnarray}

\end{widetext}

The triple time integral can be done numerically or analytically.
In the latter case, one obtains after somewhat lengthy algebra,
the following expression for the two-particle Green's function in
the atomic limit
\begin{widetext}
\begin{eqnarray}
\label{G2a}
&&  G^{2}_{\a\b}(i\omega)= \sum_{0,1,2,3,\g\d}\\
&&
(F^\a)_{01}(F^{\g\dagger})_{12}(F^\d)_{23}(F^{\b\dagger})_{30}\left[
\frac{\cR(E_1,E_2)}{E_{13}}\frac{1}{i\om-E_{10}}+
\frac{\cR(E_3,E_2)}{E_{31}}\frac{1}{i\om-E_{30}}+
\frac{\cQ(i\om,E_0,E_2)}{(i\om-E_{30})(i\om-E_{10})}
\right]\nonumber\\
&+&
(F^\a)_{01}(F^{\d})_{12}(F^{\g\dagger})_{23}(F^{\b\dagger})_{30}\left[
\frac{\cR(E_2,E_1)}{E_{13}}\frac{1}{i\om-E_{10}}+
\frac{\cR(E_2,E_3)}{E_{31}}\frac{1}{i\om-E_{30}}-
\frac{\cQ(-i\om,E_2,E_0)}{(i\om-E_{30})(i\om-E_{10})}
\right]\nonumber\\
&+&
(F^{\g\dagger})_{01}(F^{\d})_{12}(F^{\a})_{23}(F^{\b\dagger})_{30}\left[
\frac{\cR(E_2,E_1)}{E_{20}}\frac{1}{i\om-E_{32}}+
\frac{\cR(E_0,E_1)}{E_{02}}\frac{1}{i\om-E_{30}}-
\frac{\cQ(-i\om,E_3,E_1)}{(i\om-E_{30})(i\om-E_{32})}
\right]\nonumber\\
&+&
(F^{\d})_{01}(F^{\g\dagger})_{12}(F^{\a})_{23}(F^{\b\dagger})_{30}\left[
\frac{\cR(E_1,E_2)}{E_{20}}\frac{1}{i\om-E_{32}}+
\frac{\cR(E_1,E_0)}{E_{02}}\frac{1}{i\om-E_{30}}+
\frac{\cQ(i\om,E_1,E_3)}{(i\om-E_{30})(i\om-E_{32})}
\right]\nonumber\\
&+&
(F^{\d})_{01}(F^{\a})_{12}(F^{\g\dagger})_{23}(F^{\b\dagger})_{30}
\frac{1}{(i\om-E_{21})(i\om-E_{30})}\left[
\cR(E_2,E_3)-\cR(E_1,E_0)+\cQ(i\om,E_1,E_3)-\cQ(-i\omega,E_2,E_0)
\right]\nonumber\\
&+&
(F^{\g\dagger})_{01}(F^{\a})_{12}(F^{\d})_{23}(F^{\b\dagger})_{30}
\frac{1}{(i\om-E_{21})(i\om-E_{30})}\left[
\cR(E_3,E_2)-\cR(E_0,E_1)+\cQ(i\om,E_0,E_2)-\cQ(-i\omega,E_3,E_1)
\right]\nonumber
\end{eqnarray}
\end{widetext}

where we used the notation $E_{ij}=E_i-E_j$ and the functions
$\cR$ and $\cQ$ are
\begin{eqnarray}
\cR(E_1,E_2) &\equiv& (X_1+X_2)\;
T\sum_{i\om'}\frac{\Delta_{\g\d}(i\om')}{i\om'-E_{12}}\\
\cQ(i\om,E_1,E_2) &\equiv& (X_1-X_2)\;
T\sum_{i\om'}\frac{\Delta_{\g\d}(i\om')}{i\om'-i\om-E_{12}}\nonumber
\end{eqnarray}

with the atomic probabilities given by $X_i = {e^{-\beta
E_i}}/{Z}$. On the real axis these two functions take the form
\begin{eqnarray}
\cR(E_1,E_2) &\equiv& \mathrm{Re}\{X_1\;\Delta_{\g\d}^-(E_{12})-X_2\;\Delta_{\g\d}^+(E_{12})\}\\
\cQ(i\omega,E_1,E_2) &\equiv&
X_1\;\Delta_{\g\d}^-(i\om+E_{12})+X_2\;\Delta_{\g\d}^+(i\om+E_{12})\nonumber
\end{eqnarray}
where
\begin{eqnarray}
\Delta^+(z) &=& -\frac{1}{\pi}\int\frac{f(\xi)\Delta{''}(\xi)d\xi}{z-\xi}\\
\Delta^-(z) &=&
-\frac{1}{\pi}\int\frac{f(-\xi)\Delta{''}(\xi)d\xi}{z-\xi}.
\end{eqnarray}

The constant in the first term of Eq.~(\ref{G2_one}) can also be
expressed by the above defined functions
\begin{equation}
\mathrm{Tr}[G^{(atom)}\Delta] = \sum_{0,1,\d\g}
(F^\d)_{01}(F^{\g\dagger})_{10}\cR(E_1,E_0). \label{constant}
\end{equation}

Inserting Eq.~(\ref{G2a}), (\ref{constant}), (\ref{G2_one}), and
(\ref{Gatom}) into Eq.~(\ref{Sigma_exp}), we finally obtain the
self energy for the impurity models requiring the impurity levels
$\hat{t}$, the interaction matrix $\hat{U}$ and hybridization
function $\hat{\Delta}$ as an input. Thus we find a way to
calculate the exact second order perturbation in the hybridization
term. Implemented with the DMFT self consistent condition
described in reference \onlinecite{DMFT_review}, this method can
be used as a very efficient impurity solver in the DMFT study of
the multi-orbital systems with very complicated local
interactions.

\section{Benchmark}

\begin{figure}[tbp]
\begin{center}
\includegraphics[width=7cm,angle=0,clip=]{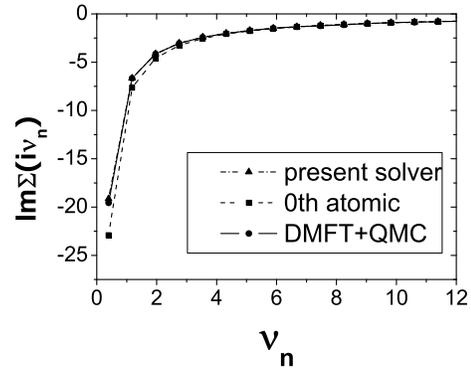}
\end{center}
\caption{The comparison of QMC and DMFT with atomic solver for two
band Hubbard model with $U=6$, $\mu=0$ ,$T=0.125$ and half band
width $D=1$.} \label{Fig.1}
\end{figure}

\begin{figure}[tbp]
\begin{center}
\includegraphics[width=7cm,angle=0,clip=]{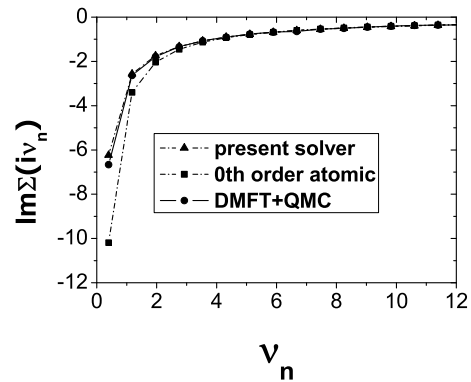}
\end{center}
\caption{The comparison of QMC and DMFT with atomic solver for two
band Hubbard model with $U=4$, $\mu=0$ ,$T=0.125$ and half band
width $D=1$.} \label{Fig.2}
\end{figure}

To test this impurity solver, we calculated the Green's function
for the two band Hubbard model at half filling and compared with
the results obtained by the QMC solver. We choose the temperature
to be $0.125$, where the QMC result is quite reliable. We use the
semi-circle density of states (DOS) and set the half bandwidth
$D=1$ as the unit of energy. The metal-insulator transition has
been determined by QMC at $U_c=3.5$. First let's compare the
results for $U=6$, where the system is on the insulator side. We
found that for frequency higher than $\omega=U$, which is the
highest energy scale in the problem, all the results obtained from
three different scheme (QMC, 0-th order atomic expansion, in which
we simply use the atomic self energy defined by Eq.~(\ref{atom})
and the present solver) fall onto a single curve, which indicates
that both the 0-th order atomic expansion and the present solver
can capture the correct high energy limit. While for the frequency
lower than $U$, the result obtained by 0-th order atomic scheme
show clear deviations  from the QMC data, including the the second
order strong coupling perturbation correction implemented by the
present solver gives excellent results  as shown in
Fig.~\ref{Fig.1}. The situation is similar for $U=4$ which is
close to the Mott transition point but still on the insulator
side. In this case, the deviation between the 0-th order result
and the QMC result becomes quite large at low frequency, as shown
in Fig.~\ref{Fig.2}. For example, the relative error reaches
$53\%$ at the first Matsubara frequency.  Again the error is
corrected by turning on the second order perturbation around the
atomic limit in the present solver. Based on the above comparison,
we can draw the conclusion that the second order perturbation in
the hybridization term works very well in the Mott insulator
phase.

\begin{figure}[tbp]
\begin{center}
\includegraphics[width=7cm,clip=]{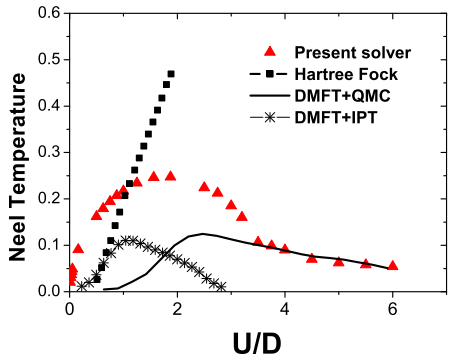}
\end{center}
\caption{The comparison of the Neel Temperature obtained by
QMC,IPT, Hatree-Fock and atomic solver for half filled single band
Hubbard model with half band width $D=1$.} \label{Neel}
\end{figure}

Another good test for this impurity solver is the
anti-ferromagnetic(AF) order in the half filled single band
Hubbard model on a bipartite lattice. In the large $U$ limit, it
is well known that in such a case up to the second order
perturbation in $t/U$, the Hubbard model can be mapped to an
anti-ferromagnetic Heisenberg model with exchange energy
$J\propto{t^2}/U$. Since the exchange energy is the only energy
scale in the problem, the Neel temperature must also be
proportional to $1/U$ in large $U$ limit. Although the physics
behind is quite straightforward, it is not reproduced by EOM or
IPT methods within the framework of DMFT. Although the
Hartree-Fock approximation can also obtain the correct AF order in
the ground state, it predicts the wrong Neel temperature to be of
order $U$ instead of $J$, as shown in Fig.~\ref{Neel}. Since the
impurity solver we proposed here can include exactly the second
order correction to the atomic limit, we expect that it can
reproduce the correct Neel temperature in the large $U$ limit. For
this purpose, we solved the single band Hubbard model at half
filling on the unfrustrated Bethe lattice and compared the results
with the Hartree-Fock approximation, DMFT+IPT and DMFT+QMC in
Fig.\ref{Neel}. Since the self energy obtained by the present
impurity solver vanishes when $U=0$, we can also get the correct
result for the non-interacting case. That is why the Neel
temperature obtained by the present solver goes down in the small
U limit. A very good agreement between our results and the QMC
results is found for $U/D>U_c$, where $U_c=3.5$ is approximately
the $U_{c2}$ in the Mott transition in the paramagnetic
phase\cite{MIT}. We also show the results obtained by iterative
perturbation theory and Hartree-Fock approximation in figure
\ref{Neel}, which are far away from the QMC results for almost the
whole range of $U$. The remarkably good agreement between our
results and the QMC data in the large $U$ limit indicates that
combined with LDA, this simple impurity solver can be used to
carry out the first principle calculation of the ordering
temperature of the materials which have spin/orbital long range
order in the ground states in the framework of LDA+DMFT. Compared
with the model Hamiltonian studies, in which the super exchange
processes are considered by the Heisenberg model, the LDA+DMFT
approach has two advantages. The first one is that unlike the
Heisenberg model, which can only capture the low energy physics,
the LDA+DMFT can capture not only the low energy physics like the
long range spin/orbital order but also the high energy physics
like the Hubbard bands. Besides that, since the effective bath in
DMFT does not come only from the nearest-neighbor sites, the
LDA+DMFT calculation can include the long range coupling between
the local spins in a natural way.

For the paramagnetic phase of an SU(N) Anderson impurity model,
the self-energy of the present solver is simple enough to be
written as a closet expression. It becomes particularly simple in
the case of half-filing where it takes the form

\begin{equation}
\Sigma(z)=\left(\frac{U}{2}\right)^2\frac{1}{z} \left(1 +
\frac{3\Delta(z)}{z}\right).
\end{equation}

To get the DMFT solution, we also need the DMFT self-consistency
condition. For the Bethe lattice, it is simply given by
$\Delta=t^2 G$ therefore the DMFT local Green's function takes the
form
\begin{equation}
G_{DMFT}(z)=\frac{1}{2\tau(z)^2}(x-s\sqrt{x^2-4\tau(z)^2})
\label{DMFT_G}
\end{equation}
where $\tau(z)=t\sqrt{1+\frac{3}{4}\frac{U^2}{z^2}}$, $x =
\frac{(z-U/2)(z+U/2)}{z}$ and
$s=\mathrm{sign}(\mathrm{Im}(x^2-4\tau(z)^2))$.

\begin{figure}
\begin{center}
\includegraphics[width=0.8\linewidth,clip=]{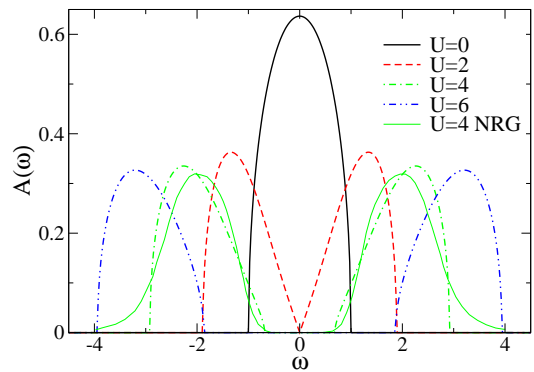}
\end{center}
\caption{ The spectral function of the one band Hubbard model on
the Bethe lattice for various values of $U$ compared with the NRG
results (taken from Ref. \onlinecite{Bulla}). } \label{Sf1}
\end{figure}

The spectral function that corresponds to Eq.~(\ref{DMFT_G}) is
plotted in Fig.~\ref{Sf1} for various values of $U$, ranging from
$U=0$ to $U=6$. For comparison, we also displayed the NRG results
at $U=4$. It is clear, that the present solver misses the Kondo
effect and therefore should be used only in the insulating state,
i.e., when the spectral function has a finite gap. At $U=4$ we can
see that the width of the gap as well as the position of the
Hubbard bands is very close to the NRG results. Note, however,
that the NRG has a finite resolution at high frequencies and
therefore does not provide very precise Hubbard bands either. They
usually tend to be slightly rounded.

It is well known that deep in the insulating state, the width of
the Hubbard bands as well as the shape of the Hubbard bands has to
be the same as the non-interacting density of states.  As one can
see in Fig.~\ref{Sf1}, the width is indeed $2D$ and they become
more and more semicircular as $U$ increases. Note that the zeroth
order approximation (expressed by Eq.~(\ref{atom})) gives for
factor of $2$ narrower Hubbard bands. Finally, the critical $U$,
at which the gap closes, is $2\sqrt{3}$ and is reasonable close to
the exact upper critical $U$ being $U_{c2}\sim 2.97$.

It should be noted that the perturbation theory in the
hybridization is a singular perturbation for the metallic system,
therefore any finite order perturbation can not offer a
qualitatively proper description of the system. If one anyway
applies the present solver to the metallic system, the local
Green's function develops a pole in the complex plane which does
not lie on the real axis. This pole indicates a tendency toward
the formation of a singularity at the Fermi level. Namely, some
weight is missing under the Hubbard bands and the spectral
function develops a V shaped cusp at the Fermi level.

\begin{figure}
\begin{center}
\includegraphics[width=0.8\linewidth,clip=]{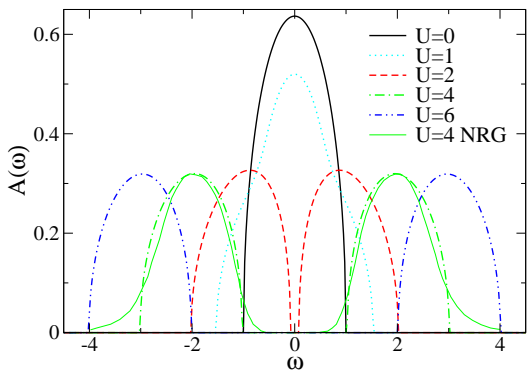}
\end{center}
\caption{ The DMFT spectral function obtained by using self-energy
from Eq.~(\ref{correctedS}) compared with the NRG results. }
\label{Sf2}
\end{figure}

To avoid the causality problem in the metallic state, one might
rewrite the self-energy in the continued fraction representation
which has the same lowest order term in expansion in $\Delta$. For
the half-filled one band model, the following self-energy can be
constructed in this way
\begin{equation}
\Sigma(z)=\left(\frac{U}{2}\right)^2\frac{1}{z-3\Delta(z)}.
\label{correctedS}
\end{equation}

The DMFT spectral function that corresponds to this self-energy is
plotted in Fig.~\ref{Sf2}. In this approximation, the system is
metallic for $U<\sqrt{3}$, however, the metallic state is not
Fermi-liquid and therefore the spectral function does not reach
the unitary limit at $\omega=0$. However, the causality problem is
avoided and the impurity solver does not break down in the
metallic state. The agreement in position and shape of the Hubbard
bands is also improved in the insulating state of the system. Note
almost perfect agreement between NRG and the present solver
spectral function at $U=4$. As an efficient impurity solver in the
strong coupling limit, the present solver should be compared with
two other impurity solvers, which are commonly used in this limit,
namely the equation of motion method(EOM)\cite{EOM} and
NCA\cite{NCA}. Unlike the method proposed in this paper, the EOM
method requires a self consistent procedure to obtain the Green's
function on the impurity site.  A general impurity model generated
by LDA+DMFT usually contains a large numbers of orbits and very
complicated non-SU(N) like local interaction. Therefore in the EOM
method one has to solve the self consistent equations with very
large number of parameters, which is almost impossible numerically
when the orbital number reaches $14$ like in systems with one open
$f$ shell per unit-cell. The NCA method suffered from the same
problem when the orbital number becomes large and the system is
away from the SU(N) symmetry since the number of atomic states,
and therefore pseudo-particles, grows exponentially. Compared with
other impurity solvers, the main advantage of the present solver
is that no self consistent loop is required to solve the impurity
problem and the local interaction can be very general including
local Coulomb repulsion, Hunds coupling, spin-orbital coupling and
pair hopping term. The computational time of the present impurity
solver still grows as $N^4$ where $N$ is the number of atomic
states, since one needs to carry out four sums in Eq.~(\ref{G2a}).
Note, however, that the matrix $F^\a_{nn'}$ has many zero elements
and if one takes into account the conservation of spin and
particle number, the computational time can be considerably
reduced.  At present, for a general $f$ system, it takes only few
seconds on modern computers.

\section{Conclusions}

To summarize, this paper presents a new impurity solver based on
second order perturbation theory in the hybridization around the
atomic limit. The  strength of the approach lies   in its
generality and  its speed. It can be applied to systems with very
complicated atomic configurations, general Coulomb repulsion and
spin-orbit coupling and does not require self-consistency
therefore it can be efficiently applied to a systems with open $d$
or $f$ shells. It is however limited to integer filling and large
$U$, which is the regime particular important to study
transition-metal compounds in the Mott insulating state with or
without orbital or magnetic long-range order.

ACKNOWLEDGEMENT:
This work was supported by the NSF under grant DMR-0096462.











\end{document}